\documentclass[aps,prc,twocolumn,superscriptaddress,showpacs,floatfix,nofootinbib]{revtex4-1}
\usepackage{amsmath,graphicx}
\begin{document}

\title{Linear and cubic response to the initial eccentricity in heavy-ion collisions}

\author{Jacquelyn Noronha-Hostler}
\affiliation{Department of Physics, Columbia University, New York, 10027, USA}
\author{Li Yan}
\affiliation{
Institut de physique th\'eorique, Universit\'e Paris Saclay, CNRS, CEA, F-91191 Gif-sur-Yvette, France} 
\author{Fernando G. Gardim}
\affiliation{Instituto de Ci\^encia e Tecnologia, Universidade Federal de Alfenas, Cidade Universit\'aria, 37715-400 Po\c cos de Caldas, MG, Brazil}
\author{Jean-Yves Ollitrault}
\affiliation{
Institut de physique th\'eorique, Universit\'e Paris Saclay, CNRS, CEA, F-91191 Gif-sur-Yvette, France} 
\date{\today}

\begin{abstract}
We study the relation between elliptic flow, $v_2$ and the initial eccentricity,  $\varepsilon_2$, in heavy-ion collisions, using hydrodynamic simulations. 
Significant deviations from linear eccentricity scaling are seen in
more peripheral collisions. 
We identify the mechanism responsible for these deviations as a cubic
response, which we argue is a generic property of the
hydrodynamic response to the initial density profile. 
The cubic response increases elliptic flow fluctuations, thereby
improving agreement of initial condition models with experimental data. 
%No significant cubic response is found for triangular flow. 
\end{abstract}


\maketitle

\section{Introduction}
Anisotropic flow, $v_n$, in heavy-ion collisions is understood as the hydrodynamic response to the anisotropy of the initial density profile. 
In hydrodynamics, $v_n$ is typically a functional of the initial density 
profile~\cite{Teaney:2010vd,Floerchinger:2013rya}.
For a given colliding system, energy, and centrality class, where the
initial density profile fluctuates event to event, one can construct in every event
predictors of elliptic flow, $v_2$, and triangular flow, $v_3$ 
 using the initial anisotropies in the corresponding
harmonics, $\varepsilon_2$ and 
$\varepsilon_3$~\cite{Alver:2006wh,Alver:2010gr,Teaney:2010vd}.  
To a good approximation, $v_2$ and $v_3$ are determined by linear
response to $\varepsilon_2$ and
$\varepsilon_3$~\cite{Gardim:2011xv,Niemi:2012aj,Gardim:2014tya,Plumari:2015cfa,Fu:2015wba}. 

Deviations from linear scaling of $v_2$ are however 
seen. In ideal hydrodynamics with a smooth, density profile,
$v_2/\varepsilon_2$  increases slightly 
for peripheral collisions~\cite{Bhalerao:2005mm}.
With a fluctuating initial density profile, the distribution of $v_2$
 differs from the distribution of $\varepsilon_2$ for
Pb-Pb collisions above 35\% centrality~\cite{Schenke:2013aza}. 
This has been recently shown to result from a slight upward curvature
of the relation between $v_2$ and $\varepsilon_2$~\cite{Niemi:2015qia}.  

In Sec.~\ref{s:kappa}, 
we show that these deviations can be quantified by adding a cubic
response term to the usual linear response. We study the variation of
the response coefficients as a 
function of centrality in hydrodynamics. 
In Sec.~\ref{s:fluctuations}, we study the effect of the cubic
response on elliptic flow fluctuations in relation with LHC 
data. 
In Sec.~\ref{s:residual}, we study the deviations
between anisotropic flow and the predictor.

\section{Linear and cubic response}
\label{s:kappa}

In a hydrodynamic calculation of a relativistic heavy ion collision, particles are emitted independently from a fluid element, 
and all information is thus contained in the single-particle momentum distribution.
This momentum distribution is determined by the initial conditions
of the hydrodynamic evolution, that is, the initial energy density profile and 
the initial fluid velocity profile. 
% also initial shear stress tensor. Is it important?
The fluid velocity at early times is itself mostly determined by the energy density profile 
at earlier times~\cite{Habich:2014jna,Liu:2015nwa}, as shown by direct inspection of hydrodynamic 
equations~\cite{Vredevoogd:2008id} and strong coupling 
calculations~\cite{vanderSchee:2012qj,Romatschke:2013re,vanderSchee:2013pia}, 
so that all observables are to a very good approximation functionals of the sole 
initial density profile. 

Anisotropic flow, $v_n$, is defined as the complex Fourier coefficient
of the single-particle azimuthal distribution in an event, that is,  
$v_n\equiv\{ e^{in\phi}\}$~\cite{Luzum:2011mm} 
where $\{\cdots\}$ denotes an average over the freeze-out 
surface~\cite{Gale:2013da} of the fluid in a single event. 
We denote by $\varepsilon_n$ the complex anisotropy in harmonic 
$n$~\cite{Bhalerao:2011yg}, defined as 
\begin{equation}
\label{defepsilon}
%\varepsilon_{n}\equiv\frac{|\{r^n e^{in\phi}\}|}{\{r^n\}},
\varepsilon_{n}\equiv -\frac{\int r^n e^{in\phi}\epsilon(r,\phi)r{\rm d}r{\rm d}\phi}
{\int r^n\epsilon(r,\phi)r{\rm d}r{\rm d}\phi},
\end{equation}
where integration is over the transverse plane 
in polar coordinates, and $\epsilon(r,\phi)$ denotes the initial energy
density at midrapidity. 
Note that the coordinate system must be centered, so that 
$\int re^{i\phi}\epsilon(r,\phi)r{\rm d}r{\rm d}\phi=0$ in every event. 
Our study in this paper is restricted to
the largest flow harmonics $n=2,3$. 
Other harmonics ($v_1$, $v_4$ and $v_5$) involve mode mixing through 
large nonlinear terms, which are already well
understood~\cite{Teaney:2012ke,Yan:2015jma}.   

We write for a given initial geometry
\begin{equation}
\label{estimator}
v_n=f(\varepsilon_n)+\delta_n,
\end{equation}
where $f(\varepsilon_n)$ is an estimator of $v_n$ based on the initial
anisotropy $\varepsilon_n$, and $\delta_n$ is the residual, defined as the difference
between the flow and the estimator. 
The estimator typically depends on a number of parameters (response
coefficients). These parameters are  
fitted in order to minimize 
$\langle |\delta_n|^2\rangle$,
where angular brackets denote averages over events in a centrality
class. 
Note that $\delta_n=0$ only if the estimator reproduces both 
the magnitude and phase of $v_n$~\cite{Gardim:2011xv}. 
In this respect, our procedure differs technically from that of 
Ref.~\cite{Niemi:2012aj}, which only retains the information on the
flow magnitude.

The eccentricity $\varepsilon_n$ in a given harmonic 
transforms like $v_n$ under azimuthal rotations. 
Therefore the estimator $f(\varepsilon_n)$ must also transform like
$\varepsilon_n$ under azimuthal rotations. 
The simplest choice is 
\begin{equation}
\label{linear}
f(\varepsilon_n)=\kappa_n\varepsilon_n,
\end{equation}
corresponds to linear eccentricity
scaling~\cite{Teaney:2010vd,Gardim:2011xv}.
The lowest nonlinear correction preserving rotational symmetry and
analyticity is a cubic response term~\cite{Gardim:2014tya,Yan:2014nsa}: 
\begin{equation}
\label{cubic}
f(\varepsilon_n)=\kappa_n\varepsilon_n+\kappa'_n|\varepsilon_n|^2\varepsilon_n ,
\end{equation}
where $\kappa_n$ is the linear response coefficient and 
$\kappa'_n$ the cubic response coefficient. 
Parity requires that $\kappa_n$ and $\kappa'_n$ are both real.  
Their explicit expressions are derived in Appendix~\ref{s:appendix}. 
Note that the values of $\kappa_n$ in Eqs.~(\ref{linear}) and
(\ref{cubic}) differ in general, i.e., the linear response coefficient
is modified by the cubic response. 

\begin{figure}[h]
\begin{center}
\includegraphics[width=\linewidth]{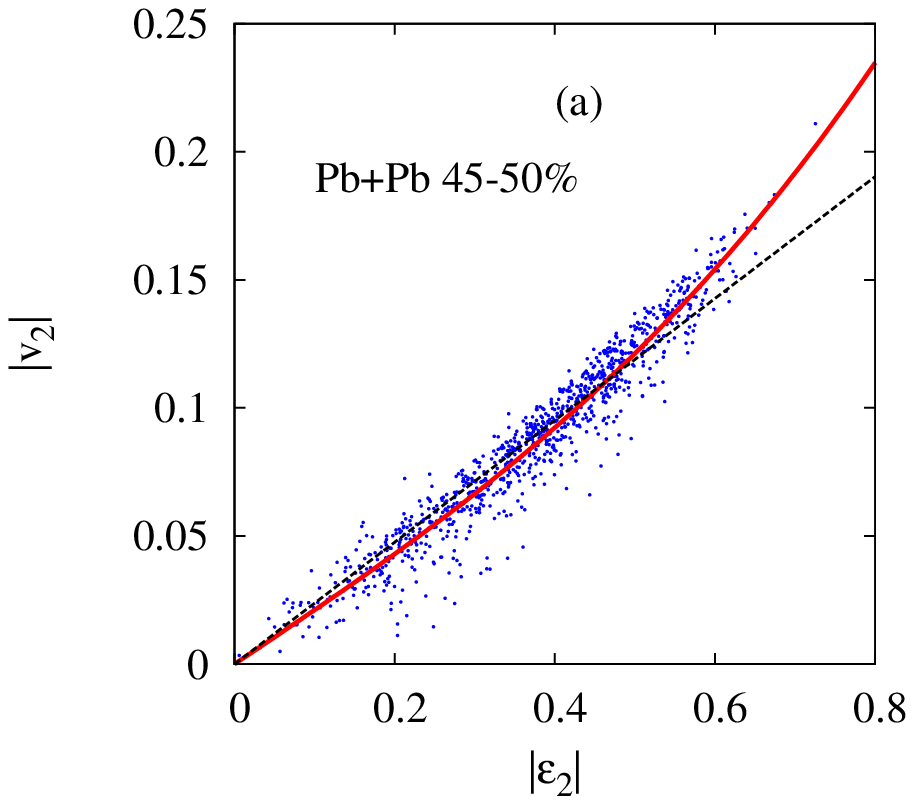} 
\includegraphics[width=\linewidth]{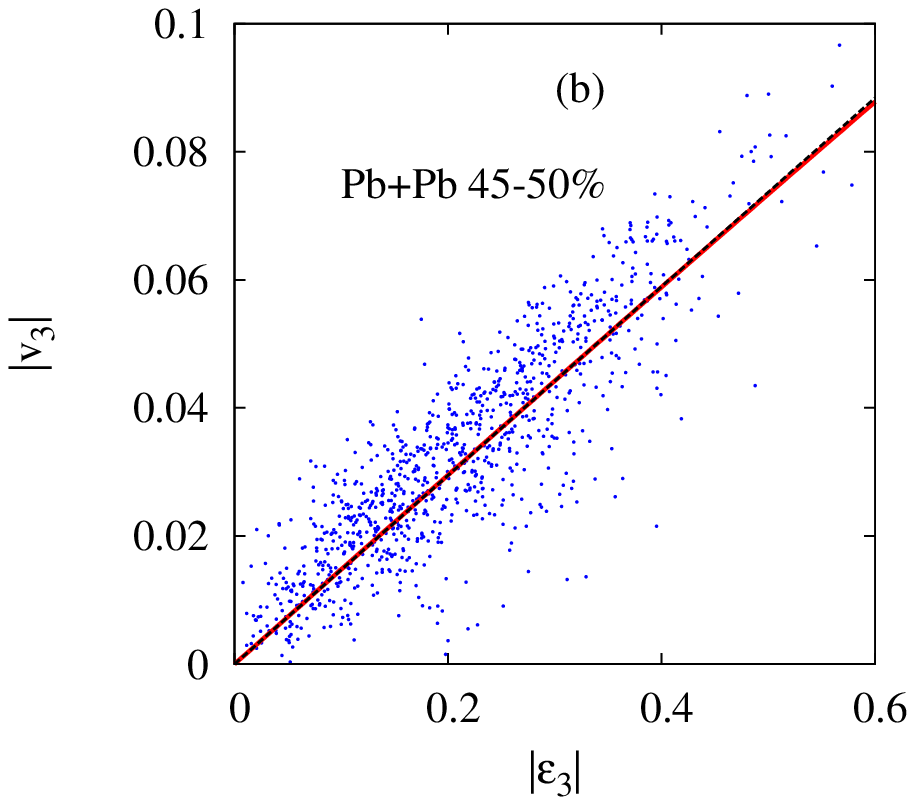} 
\end{center}
\caption{(Color online) Correlation between the magnitudes of
 anisotropic flow $v_n$ and
  initial eccentricity $\varepsilon_n$ 
for Pb+Pb collisions at $2.76$~TeV in the 45-50\% centrality range. 
Each point corresponds to a different initial geometry. 
Dotted line: linear estimator, Eq.~(\ref{linear}).
Full line: cubic estimator, Eq.~(\ref{cubic}).
(a) Elliptic flow. (b) Triangular flow. 
\label{fig:scatter}
}
\end{figure} 

We calculate $v_n$ using the boost-invariant~\cite{Bjorken:1982qr} 2+1 dimensional viscous relativistic hydrodynamical code v-USPhydro 
\cite{Noronha-Hostler:2013gga,Noronha-Hostler:2014dqa}.
The initial conditions are calculated using a Monte Carlo Glauber
model \cite{Miller:2007ri,Alver:2008aq,Rybczynski:2013yba}  
for Pb+Pb collisions at $\sqrt{s_{NN}}=2.76$~TeV. 
The energy density at an initial time $\tau_0=0.6$~fm/c after the
collision is assumed to be proportional to the 
density of binary collisions~\cite{Hirano:2009ah}.
%we need to define this quantity better 
The centrality of the event is defined according to the number of participant nucleons~\cite{Noronha-Hostler:2014dqa}. 
For each 5\% centrality class, we generate approximately 1000 events.   
We assume for simplicity that there is no initial transverse flow velocity $u^{x} = u^y =0$, and 
that the bulk pressure, $\Pi$, and the shear stress tensor, $\pi^{\mu\nu}$, vanish at $\tau_0$.  
We use a constant shear viscosity over entropy ratio $\eta/s=1/4\pi$~\cite{Policastro:2001yc,Kovtun:2004de},
and zero bulk viscosity.   While a temperature dependent $\eta/s(T)$ and $\zeta/s(T)$ may be more realistic such as from \cite{NoronhaHostler:2008ju,Niemi:2011ix}, it is unlikely that either would have a large impact on the results because the mapping is nearly identical between our constant $\eta/s$ and the $\eta/s(T)+\zeta/s(T)$ from \cite{Gardim:2014tya}.  
The equation of state is that of Ref.~\cite{Huovinen:2009yb} with vanishing baryon chemical potential. 
We have adopted  the popular quadratic ansatz for the viscous correction to the thermal distribution 
function~\cite{Teaney:2003kp,Dusling:2009df} and a constant freeze-out temperature $T_{FO}=130$~MeV.
We calculate $v_n$ 
for pions emitted directly at freeze-out over the transverse momentum range $0.3<p_t<3$~GeV/c~\cite{Chatrchyan:2012ta}. 

Fig.~\ref{fig:scatter} displays scatter plots of the magnitudes of
initial anisotropies $|\varepsilon_n|$ and anisotropic flow $|v_n|$
for $n=2,3$ in Pb+Pb collisions at 2.76~TeV in the 45-50\% centrality
range. 
The linear and cubic estimators (\ref{linear}) and (\ref{cubic}) are
also shown as dashed and solid lines, respectively. 
Note that these lines do not strictly correspond to best fits of the set of
points: the magnitude of the best fit does not coincide with 
the best fit to the magnitudes (it is slightly lower), 
because the optimization of the estimator also involves the phases 
(see Appendix~\ref{s:appendix} for details). 
For elliptic flow, a clear departure from linear scaling is seen for 
 large $|\varepsilon_2|$~\cite{Niemi:2015qia}, which is captured 
 by the cubic term, and corresponds to a positive $\kappa'_2$. 
For triangular flow, such nonlinear effects are negligible. 
The dispersion of the results around the best-fit curve is studied in 
Sec.~\ref{s:residual}.

\begin{figure}[h]
\begin{center}
\includegraphics[width=\linewidth]{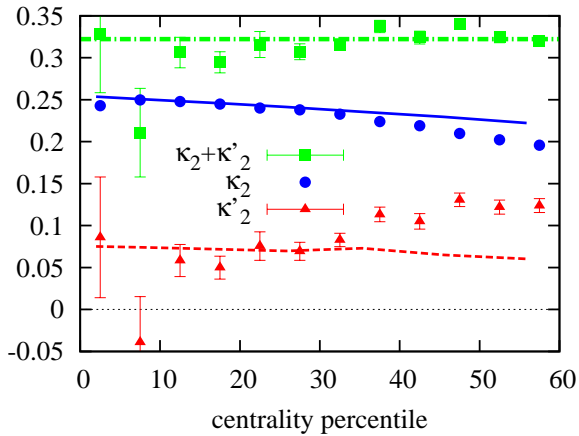} 
\end{center}
\caption{(Color online) Response coefficients, as defined by
  Eqs.~(\ref{cubic}), %for elliptic flow, 
for Pb+Pb collisions at  $\sqrt{s_{NN}}=2.76$~TeV, 
as a function of centrality percentile. 
  Circles and solid line: linear response coefficient $\kappa_2$ with fluctuating and smooth initial conditions.   
  Triangles and dashed line: cubic response coefficient $\kappa'_2$ with fluctuating and smooth initial conditions.
 Squares: $\kappa_2+\kappa'_2$.  The dash-dotted horizontal line illustrates that  $\kappa_2+\kappa'_2\simeq 0.32$  for all centralities. 
 \label{fig:response}
}
\end{figure}    
The values of $\kappa_2$ and $\kappa'_2$ from Eq.~(\ref{cubic}) 
are displayed in Fig.~\ref{fig:response} as a function of centrality. 
Statistical errors due to the finite number of
events, shown as vertical bars in figures, are estimated by 
jackknife resampling~\cite{jackknife}.
A cubic response clearly appears above $10\%$ centrality, although 
it is too small to be seen by visual inspection of the scatter plots 
below 40\% centrality. 
While the linear response decreases with centrality, as expected as a consequence of 
viscous suppression~\cite{Drescher:2007cd}, the cubic response {\it increases\/} with centrality, 
in such a way that the sum $\kappa_2+\kappa'_2$ remains approximately constant 
(squares in Fig.~\ref{fig:response}).
For sake of comparison, we have also carried out ideal hydrodynamic 
simulations for selected centrality bins (not shown).
The linear response coefficient is larger than for viscous hydrodynamics, as expected, but 
the cubic coefficient is smaller: $\kappa_2$ and $\kappa'_2$ again vary in 
opposite directions. 
We have also carried out calculations with
MCKLN~\cite{Drescher:2006ca} initial conditions (not shown) and
compared with the results from Glauber initial conditions. 
We find that the variations of $\kappa_2$ and $\kappa'_2$ as a
function of centrality are similar, 
but stronger: in particular, $\kappa_2$ becomes smaller than
$\kappa'_2$ for the most peripheral bin. 
The sum $\kappa_2+\kappa'_2$ is also approximately constant. 

In Eq.~(\ref{estimator}), one expects that the estimator $f(\varepsilon_2)$ 
captures the long-range structure of the initial density profile~\cite{Teaney:2010vd,Bhalerao:2011bp,Blaizot:2014nia}
while the residual $\delta_2$ is driven by short-range structures. 
In order to test this hypothesis, we repeat the calculation with a smooth Gaussian initial 
density profile, for which one expects $\delta_2\approx 0$.\footnote{In particular, the phase of 
$\delta_2$ is exactly zero due to symmetry of the Gaussian.} 
In every centrality bin, we fix the total  entropy and rms radius to the same value as in the 
previous calculation with fluctuating initial conditions. 
We calculate $v_2$ for two different values of $\varepsilon_2$ ($0.15$
and $0.25$) and determine 
$\kappa_2$ and $\kappa'_2$ by solving 
$v_2=\kappa_2\varepsilon_2+\kappa'_2\varepsilon_2^3$. We use a third
value of $\varepsilon_2$ ($0.3$) to check that results are compatible.

Our calculation with smooth initial conditions uses a different code~\cite{Teaney:2012ke}
than the calculation with fluctuating initial conditions. The differences are the following. 
The shear tensor $\Pi^{\mu\nu}$ is initialized to the 
Navier-Stokes value, not to 0, but this is known to have a negligible effect at late times~\cite{Luzum:2008cw}. 
The equation of state is that of Ref.~\cite{Laine:2006cp}, but we have also checked that this has 
a negligible effect. 
Finally, the event-by-event calculations are done with the Lagrangian
method known as Smoothed Particle Hydrodynamics~\cite{Aguiar:2000hw,Hama:2004rr} while the smooth
initial conditions are done within a grid method. However, both reproduce exact solutions \cite{Marrochio:2013wla} so the results should be comparable.

The results with smooth initial conditions are shown as lines in Fig.~\ref{fig:response}. 
Up to 30\% centrality, smooth initial conditions and fluctuating initial conditions give very similar results
for both $\kappa_2$ and $\kappa'_2$. 
Above 30\% centrality, the centrality dependence is stronger with fluctuating initial conditions than 
with smooth initial conditions. In particular, no increase of $\kappa'_2$ with centrality percentile is observed 
with smooth initial conditions. 
This difference between smooth initial conditions and fluctuating initial conditions appears --- as it should ---  when the size of the system is smaller and becomes comparable to the size of the fluctuations. 
We also find (not shown in figure) that the cubic response coefficient is slightly larger in ideal hydrodynamics 
than in viscous hydrodynamics, while the opposite variation is seen with fluctuating initial conditions. 

Thus all our hydrodynamic calculations, ideal or viscous, with or without fluctuations confirm that a cubic response exists in addition to the well known linear response. The effect of the cubic response is negligible for central collisions but becomes sizable as the centrality percentile increases. Around 50\%  centrality, about 10\%  of the elliptic flow comes from the cubic term, hence the cubic response matters for precision studies. 

For $v_3$, a similar analysis shows 
the relevant cubic response term is proportional to 
$|\varepsilon_2|^2\varepsilon_3$, not
$|\varepsilon_3|^2\varepsilon_3$. 
Detailed results are presented in Appendix~\ref{s:v3}.

\section{Application to elliptic flow fluctuations}
\label{s:fluctuations}
\begin{figure}[h]
\begin{center}
\includegraphics[width=\linewidth]{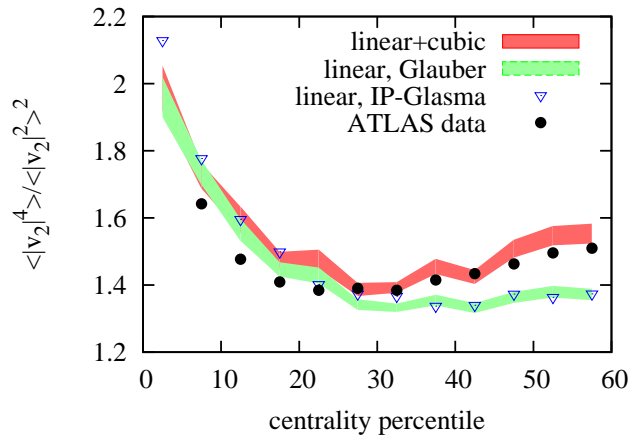} 
\end{center}
\caption{(Color online) Ratio
$\langle v_2^4\rangle/\langle v_2^2\rangle^2$ for Pb+Pb collisions at
  $\sqrt{s_{NN}}=2.76$~TeV as a function of centrality percentile. 
Light shaded band: assuming linear response, with Monte-Carlo Glauber initial conditions. 
Dark shaded band: with cubic response added. 
Triangles: assuming linear response with IP Glasma initial conditions \cite{Schenke:2012wb,Schenke:2013aza}.
Circles: ATLAS data \cite{Aad:2014vba}.
\label{fig:ratios}
}
\end{figure}    
We now discuss the effect of cubic response on elliptic flow
fluctuations. 
The magnitude of flow fluctuations can be quantified by 
ratios of cumulants~\cite{Borghini:2001vi,Bilandzic:2010jr} 
or moments~\cite{Bhalerao:2014xra} of the distribution of $v_2$.
The simplest ratio is \cite{Bhalerao:2011yg} 
$\langle |v_2|^4\rangle/\langle |v_2|^2\rangle^2$, where angular
brackets denote an average over events in a centrality class. 
Neglecting $\delta_n$ in Eq.~(\ref{estimator}),  Eq.~(\ref{cubic}) gives,
to leading order in the cubic response
$\kappa'_2$:
\begin{equation}
\label{momentratio}
\frac{\langle |v_2|^4\rangle}{\langle |v_2|^2\rangle^2}\simeq 
\frac{\langle |\varepsilon_2|^4\rangle}{\langle
  |\varepsilon_2|^2\rangle^2}
\left(
1+4\frac{\kappa'_2}{\kappa_2}
\left(
\frac{\langle |\varepsilon_2|^6\rangle}{\langle
  |\varepsilon_2|^4\rangle}
-\frac{\langle |\varepsilon_2|^4\rangle}{\langle
  |\varepsilon_2|^2\rangle}
\right)\right).
\end{equation}
The left-hand side 
differs from the right-hand side by less than $0.02$ for all centralities, which means 
that the ratio of moments of the $v_2$ distribution is determined to an excellent approximation
by the corresponding ratio of eccentricities, corrected by the cubic response. 

When $\kappa'_2>0$, the cubic response increases the ratio. 
The shaded bands in Fig.~\ref{fig:ratios} display the right-hand side of
Eq.~(\ref{momentratio}) with and without the cubic response
$\kappa'_2$, for our Monte-Carlo Glauber model of initial conditions. 
With linear response alone,  the ratio is slightly too large for central collisions 
and too low for peripheral collisions. 
The cubic response leaves the ratio unchanged for central collisions
but increases it by up to 15\% for peripheral collisions where 
it significantly improves agreement with experimental
data~\cite{Aad:2014vba}.  

In a previous hydrodynamic study using as initial condition the IP-Glasma model~\cite{Schenke:2013aza},
it was found that the distribution of $v_2$ matches experimental data for all centralities 
while the distribution of $\varepsilon_2$ is too narrow for centralities above 35\%. 
This observation is naturally explained by the cubic response. 
Figure~\ref{fig:ratios} shows that the fluctuations of $\varepsilon_2$
are very similar with the IP-Glasma model
and with the Monte-Carlo Glauber. 

The fact that linear eccentricity scaling alone underpredicts the
ratio for peripheral collisions has also been noted
previously~\cite{Bhalerao:2011yg} using Monte-Carlo Glauber and
MCKLN~\cite{Drescher:2006ca} models. 
It seems a generic feature of existing models of initial conditions. 
Once the cubic response is taken into account, one expects models to
be in better agreement with data on elliptic flow fluctuations. 

\section{Residual analysis}
\label{s:residual}

Figure~\ref{fig:scatter} shows that there is a significant dispersion of anisotropic 
flow for a given initial anisotropy, i.e., a significant residual $\delta_n$. 
For elliptic flow, the magnitude of $\delta_2$ is typically 10\% of the value of $v_2$, 
which is as large or larger than the cubic response. 
But unlike the cubic response, the residual averages to zero, so that its effect on measured
quantities, which are averaged over many events, is small. 
For instance, its contribution  to the mean square elliptic flow is 
proportional to $|\delta_2|^2$, as shown by Eq.~(\ref{delta2}). 
Therefore, the correction from the residual to the rms value of $v_2$ is typically less than 1\% 
in relative value. 
\begin{figure}
\begin{center}
\includegraphics[width=\linewidth]{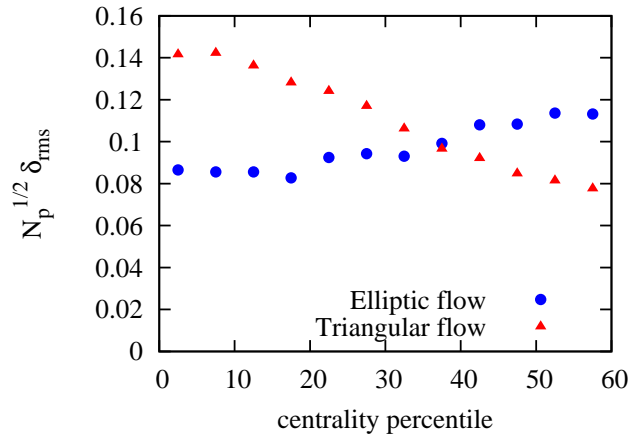} 
\end{center}
\caption{(Color online) rms value of the residual $\delta_n$ multiplied by  $N_p^{1/2}$, 
where $N_p$ is the average number of participants in the centrality bin, as a function of 
centrality percentile, for the same viscous hydrodynamic calculation as in Fig.~\ref{fig:response}.
\label{fig:residual}
}
\end{figure}    

The residual $\delta_n$ is due to short-range fluctuations whose effect is not captured by the 
eccentricity $\varepsilon_n$. 
Fluctuations in our calculation are due to the finite number of participant nucleons $N_p$. 
Therefore one naturally expects that the magnitude of $\delta_n$ scales roughly like $N_p^{-1/2}$. 
Figure~\ref{fig:residual} displays $\sqrt{N_p\langle |\delta_n|^2\rangle}$ as a function of the 
centrality percentile from our viscous hydrodynamic calculation for $n=2,3$. 
One sees that it varies by less than a factor 2, while the number of participants varies almost 
by a factor 10. 
The decrease of $\langle |\delta_3|^2\rangle/\langle |\delta_2|^2\rangle$
as a function of centrality percentile seen in Fig.~\ref{fig:residual}
can be ascribed to the larger damping of $v_3$, relative to
$v_2$~\cite{Alver:2010dn}. 

We have checked that $\langle |\delta_n|^2|f_n|^2\rangle=\langle
|\delta_n|^2\rangle\langle|f_n|^2\rangle$
within errors for
all centralities. 
This means that 
the magnitude of the residual is independent of
that of the estimator, a property referred to as {\it
  homoscedasticity\/}. 

Finally, we have studied whether the distribution of $\delta_n$ is isotropic. 
The projection of $\delta_n$ parallel to the estimator $f(\varepsilon_n)$ corresponds 
to the dispersion in the magnitude $|v_n|$, while the projection perpendicular to 
$f(\varepsilon_n)$ corresponds to the dispersion in the flow angle. 
For elliptic flow, we find a sizable anisotropy as the centrality percentile increases: 
$\langle [{\rm Re}(\delta_2^* f(\varepsilon_2))]^2\rangle >\langle [{\rm Im}(\delta_2^* f(\varepsilon_2))]^2\rangle$, 
which means that the relative fluctuations of the flow magnitude with
respect to the estimator are larger than the fluctuations of the flow angle.

\section{Conclusions}

We find that the elliptic flow not only has a contribution from the usual linear response from the initial eccentricities but there is a nonzero cubic response that plays a strong role for mid-central to peripheral collisions, which highlights the importance of cubic response for large eccentricities.   In fact, its contribution to the total elliptic flow is of order 10\% at around 50\% centrality. The existence of  non-zero cubic response indicates that the distribution of $v_2$ is not homothetic to the distribution of $\varepsilon_2$, as usually
assumed~\cite{Aad:2013xma,Retinskaya:2013gca,Renk:2014jja,Yan:2014nsa,Rybczynski:2015wva}. 
%Hadronic rescatterings may have a similar effect~\cite{Fu:2015wba}. 

Because we consistently see this effect regardless of the scale of fluctuations and  the type of viscosity, we conclude that it is a general property of the hydrodynamic response.  Current calculations are for Pb+Pb collisions at $\sqrt{s_{NN}}=2.76$~TeV, however, previous results at  Au+Au RHIC energies \cite{Gardim:2014tya} qualitatively appear to have nearly identical response.  Most likely smaller, asymmetric systems would display similar effects but they may see a larger influence from small scale structure~\cite{Noronha-Hostler:2015coa}. 

The sum of the linear and cubic response is approximately constant across centralities so one would naively expect a simple explanation.  Note that this quantity corresponds (see Eq.~(\ref{cubic})) to the
limiting value of $v_2$ for $\varepsilon_2\to 1$, i.e, to the emission 
from a one-dimensional source. 
While both smoothed and event-by-event initial conditions see a non-zero cubic response, the magnitude of the each is quite different across centralities.  With fluctuating initial conditions, the cubic response coefficient, 
$\kappa'_2$ consistently increases as a function of centrality percentile whereas it is roughly constant for smoothed initial conditions. Thus, for event-by-event initial conditions the cubic response is large precisely in the region where the cubic response is most relevant.  Conversely, the linear response coefficient, $\kappa_2$, decreases across centralities at a steeper rate for event-by-event fluctuations.  
We conclude then that the cubic response depends on the detailed  
structure of initial conditions with a non-trivial dependence on the small scale fluctuations of the initial
density profile, which deserves further investigations. 

\begin{acknowledgments}
JNH  acknowledges support
from the US-DOE Nuclear Science Grant No. DE-FG02-93ER40764. 
LY is funded  by the European Research Council under the 
Advanced Investigator Grant ERC-AD-267258. FGG  was supported by
Conselho Nacional de Desenvolvimento Cient\'\i fico e Tecnol\'ogico
(CNPq) No. 449694/2014-3, and Fapemig.
We thank Matt Luzum for useful discussions. 
JYO thanks the Tata Institute of Fundamental Research for hospitality
while this work was being completed. 
\end{acknowledgments}

\appendix
\section{Expressions of response coefficients}
\label{s:appendix}

If the estimator $f(\varepsilon_n)$ in Eq.~(\ref{estimator}) depends
on a number of parameters, minimizing 
$\langle |\delta_n|^2\rangle$ 
with respect to these parameters gives the condition
\begin{equation}
\label{minimization}
{\rm Re}\langle (v_n-f(\varepsilon_n))\, df^*(\varepsilon_n)\rangle=
{\rm Re}\langle \delta_n\, df^*(\varepsilon_n)\rangle=0.
\end{equation}

Differentiating with respect to $\kappa_n$ in Eq.~(\ref{linear}) or
Eq.~(\ref{cubic}) (keeping $\kappa'_n/\kappa_n$ constant) gives 
$df\propto f$, 
and Eq.~(\ref{minimization}) gives 
\begin{equation}
\label{kappa}
{\rm Re}\langle v_n f^*(\varepsilon_n)\rangle=
\langle |f(\varepsilon_n)|^2\rangle.
\end{equation}
This equation allows to relate the difference $\langle
|\delta_n|^2\rangle$ to the Pearson correlation coefficient between the
flow $v_n$ and the estimator $f(\varepsilon_n)$. 
Using Eq.~(\ref{estimator}) and Eq.~(\ref{kappa}), one obtains
\begin{equation}
\label{delta2}
\langle |\delta_n|^2\rangle=\langle |v_n|^2\rangle-\langle |f(\varepsilon_n)|^2\rangle.
\end{equation}
The Pearson correlation coefficient is defined as 
\begin{equation}
\label{pearson}
Q_n\equiv\frac{{\rm Re}\langle v_n f^*(\varepsilon_n)\rangle}
{\sqrt{\langle |v_n|^2\rangle\langle|f(\varepsilon_n)|^2\rangle}}=
\sqrt{\frac{\langle|f(\varepsilon_n)|^2\rangle}{\langle |v_n|^2\rangle}},
\end{equation}
where, in the last equality, we have used Eq.~(\ref{kappa}). 
$Q_n$ lies between $-1$ and $+1$. 
Using this equation, Eq.(\ref{delta2}) gives
\begin{equation}
\frac{\langle |\delta_n|^2\rangle}{\langle |v_n|^2\rangle}=
1-Q_n^2. 
\end{equation}
When $Q_n$ is close to 1, the difference between the flow and the
estimator is small, as expected. 

With a purely linear response, Eq.~(\ref{linear}), the expression of
the coefficient is~\cite{Gardim:2011xv}
 \begin{equation}
\label{kappal}
\kappa_n=\frac{
{\rm Re}\left(\langle  v_n\varepsilon_n^*\rangle\right)}
{\langle |\varepsilon_n|^2\rangle}.
\end{equation}
When a cubic response term is added, Eq.~(\ref{cubic}), one must
minimize $\langle|\delta_n|^2\rangle$ with respect to $\kappa_n$ and
$\kappa'_n$. This yields a system of two equations whose solution is
\begin{eqnarray}
\label{kappaprime}
\kappa_n&=&\frac{
{\rm Re}\left(
\langle |\varepsilon_n|^6\rangle\langle  v_n\varepsilon_n^*\rangle
-\langle |\varepsilon_n|^4\rangle\langle
v_n\varepsilon_n^*|\varepsilon_n|^2\rangle\right)}
{\langle |\varepsilon_n|^6\rangle\langle |\varepsilon_n|^2\rangle-\langle
  |\varepsilon_n|^4\rangle^2}\cr
\kappa'_n&=&\frac{
{\rm Re}\left(
-\langle |\varepsilon_n|^4\rangle\langle  v_n\varepsilon_n^*\rangle
+\langle |\varepsilon_n|^2\rangle\langle
v_n\varepsilon_n^*|\varepsilon_n|^2\rangle\right)}
{\langle |\varepsilon_n|^6\rangle\langle |\varepsilon_n|^2\rangle-\langle
  |\varepsilon_n|^4\rangle^2}.
\end{eqnarray}
%The denominators in the right-hand side are positive thanks to
%H\"older's inequality
Note that the expression of the linear response coefficient $\kappa_n$
is modified by including a cubic response. 
On the other hand, the cubic response only increases the Pearson
coefficient $Q_n$ by a negligible amount. 
\begin{figure}
\begin{center}
\includegraphics[width=\linewidth]{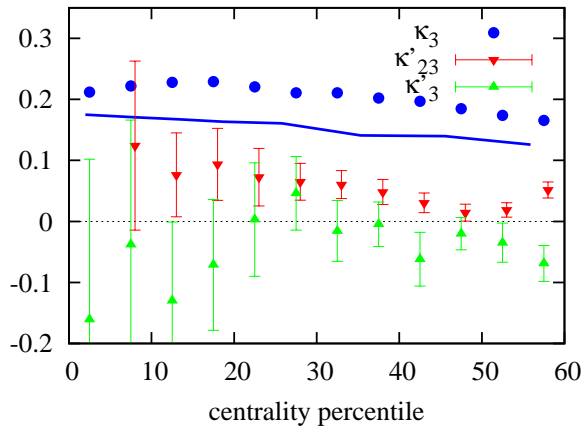} 
\end{center}
\caption{(Color online) Linear and cubic response coefficients, as
  defined by Eqs.~(\ref{cubic}), for triangular flow. 
  Solid line and full circles: $\kappa_3$ with smooth and fluctuating initial conditions. 
Triangles: $\kappa'_3$ and $\kappa'_{23}$ (see Eq.(\ref{kappa23})) with
fluctuating initial conditions. \label{fig:response3}}
\end{figure}   

\section{Triangular flow}
\label{s:v3}

We have carried out the same analysis for $v_3$ as for $v_2$. 
With fluctuating initial conditions, 
the cubic response $\kappa'_3$ defined in Eq.~(\ref{cubic}) is
compatible with zero within statistical error bars, as seen in  
Fig.~\ref{fig:response3}. Considering all centralities together, a
negative value is  preferred. 
We have also tested a different type of cubic response mixing the
second and third harmonic, namely:
\begin{equation}
\label{kappa23}
v_3=\kappa_3\varepsilon_3+\kappa'_{23}|\varepsilon_2|^2\varepsilon_3+\delta_3.
\end{equation}
Since $|\varepsilon_2|$ is significantly larger than $|\varepsilon_3|$
for mid-central collisions, one expect that such a term could be
larger than a cubic term involving just $\varepsilon_3$. 
The values of $\kappa'_{23}$ are plotted in Fig.~\ref{fig:response3}
as a function of centrality. Considering all 
centralities together, there is significant evidence for a small positive 
$\kappa'_{23}\sim 0.03$. 

Figure~\ref{fig:response3} also presents our results for the linear 
response $\kappa_3$. 
Its value is essentially the same whether or
not one includes cubic terms in the fit. 
It varies less than $\kappa_2$ as a function of centrality. 
We have also carried out a calculation with smooth initial conditions obtained by a triangular 
deformation of a symmetric  Gaussian~\cite{Teaney:2010vd}. 
The  resulting values of $\kappa_3$, shown as a solid curve in
Fig.~\ref{fig:response3}, are close to those obtained with fluctuating
initial conditions.  
%The values of $\kappa'_3$ are small (typically smaller than $0.05$)
%but fluctuate somewhat due to numerical errors; they are therefore not
%plotted.   


\begin{thebibliography}{99}
%\cite{Teaney:2010vd}
\bibitem{Teaney:2010vd} 
  D.~Teaney and L.~Yan,
  %``Triangularity and Dipole Asymmetry in Heavy Ion Collisions,''
  Phys.\ Rev.\ C {\bf 83}, 064904 (2011)
  doi:10.1103/PhysRevC.83.064904
  [arXiv:1010.1876 [nucl-th]].
  %%CITATION = doi:10.1103/PhysRevC.83.064904;%%
  %209 citations counted in INSPIRE as of 19 Jan 2016


%\cite{Floerchinger:2013rya}
\bibitem{Floerchinger:2013rya} 
  S.~Floerchinger and U.~A.~Wiedemann,
  %``Mode-by-mode fluid dynamics for relativistic heavy ion collisions,''
  Phys.\ Lett.\ B {\bf 728}, 407 (2014)
  doi:10.1016/j.physletb.2013.12.025
  [arXiv:1307.3453 [hep-ph]].
  %%CITATION = doi:10.1016/j.physletb.2013.12.025;%%
  %20 citations counted in INSPIRE as of 19 Jan 2016


%\cite{Alver:2006wh}
\bibitem{Alver:2006wh} 
  B.~Alver {\it et al.} [PHOBOS Collaboration],
  %``System size, energy, pseudorapidity, and centrality dependence of elliptic flow,''
  Phys.\ Rev.\ Lett.\  {\bf 98}, 242302 (2007)
  doi:10.1103/PhysRevLett.98.242302
  [nucl-ex/0610037].
  %%CITATION = doi:10.1103/PhysRevLett.98.242302;%%
  %256 citations counted in INSPIRE as of 19 Jan 2016


%\cite{Alver:2010gr}
\bibitem{Alver:2010gr} 
  B.~Alver and G.~Roland,
  %``Collision geometry fluctuations and triangular flow in heavy-ion collisions,''
  Phys.\ Rev.\ C {\bf 81}, 054905 (2010)
  [Phys.\ Rev.\ C {\bf 82}, 039903 (2010)]
  doi:10.1103/PhysRevC.82.039903, 10.1103/PhysRevC.81.054905
  [arXiv:1003.0194 [nucl-th]].
  %%CITATION = doi:10.1103/PhysRevC.82.039903, 10.1103/PhysRevC.81.054905;%%
  %475 citations counted in INSPIRE as of 19 Jan 2016


%\cite{Gardim:2011xv}
\bibitem{Gardim:2011xv} 
  F.~G.~Gardim, F.~Grassi, M.~Luzum and J.~Y.~Ollitrault,
  %``Mapping the hydrodynamic response to the initial geometry in heavy-ion collisions,''
  Phys.\ Rev.\ C {\bf 85}, 024908 (2012)
  doi:10.1103/PhysRevC.85.024908
  [arXiv:1111.6538 [nucl-th]].
  %%CITATION = doi:10.1103/PhysRevC.85.024908;%%
  %112 citations counted in INSPIRE as of 19 Jan 2016


%\cite{Niemi:2012aj}
\bibitem{Niemi:2012aj} 
  H.~Niemi, G.~S.~Denicol, H.~Holopainen and P.~Huovinen,
  %``Event-by-event distributions of azimuthal asymmetries in ultrarelativistic heavy-ion collisions,''
  Phys.\ Rev.\ C {\bf 87}, no. 5, 054901 (2013)
  doi:10.1103/PhysRevC.87.054901
  [arXiv:1212.1008 [nucl-th]].
  %%CITATION = doi:10.1103/PhysRevC.87.054901;%%
  %88 citations counted in INSPIRE as of 19 Jan 2016


%\cite{Gardim:2014tya}
\bibitem{Gardim:2014tya} 
  F.~G.~Gardim, J.~Noronha-Hostler, M.~Luzum and F.~Grassi,
  %``Effects of viscosity on the mapping of initial to final state in heavy ion collisions,''
  Phys.\ Rev.\ C {\bf 91}, no. 3, 034902 (2015)
  doi:10.1103/PhysRevC.91.034902
  [arXiv:1411.2574 [nucl-th]].
  %%CITATION = doi:10.1103/PhysRevC.91.034902;%%
  %13 citations counted in INSPIRE as of 19 Jan 2016


%\cite{Plumari:2015cfa}
\bibitem{Plumari:2015cfa} 
  S.~Plumari, G.~L.~Guardo, F.~Scardina and V.~Greco,
  %``Initial state fluctuations from mid-peripheral to ultra-central collisions in a event-by-event transport approach,''
  Phys.\ Rev.\ C {\bf 92}, no. 5, 054902 (2015)
  doi:10.1103/PhysRevC.92.054902
  [arXiv:1507.05540 [hep-ph]].
  %%CITATION = doi:10.1103/PhysRevC.92.054902;%%
  %3 citations counted in INSPIRE as of 19 Jan 2016


%\cite{Fu:2015wba}
\bibitem{Fu:2015wba} 
  J.~Fu,
  %``Centrality dependence of mapping the hydrodynamic response to the initial geometry in heavy-ion collisions,''
  Phys.\ Rev.\ C {\bf 92}, no. 2, 024904 (2015).
  doi:10.1103/PhysRevC.92.024904
  %%CITATION = doi:10.1103/PhysRevC.92.024904;%%
  %2 citations counted in INSPIRE as of 19 Jan 2016


%\cite{Bhalerao:2005mm}
\bibitem{Bhalerao:2005mm} 
  R.~S.~Bhalerao, J.~P.~Blaizot, N.~Borghini and J.~Y.~Ollitrault,
  %``Elliptic flow and incomplete equilibration at RHIC,''
  Phys.\ Lett.\ B {\bf 627}, 49 (2005)
  doi:10.1016/j.physletb.2005.08.131
  [nucl-th/0508009].
  %%CITATION = doi:10.1016/j.physletb.2005.08.131;%%
  %169 citations counted in INSPIRE as of 19 Jan 2016


%\cite{Schenke:2013aza}
\bibitem{Schenke:2013aza} 
  B.~Schenke, P.~Tribedy and R.~Venugopalan,
  %``Gluon field fluctuations in nuclear collisions: Multiplicity and eccentricity distributions,''
  Nucl.\ Phys.\ A {\bf 926}, 102 (2014)
  doi:10.1016/j.nuclphysa.2014.03.001
  [arXiv:1312.5588 [hep-ph]].
  %%CITATION = doi:10.1016/j.nuclphysa.2014.03.001;%%
  %10 citations counted in INSPIRE as of 19 Jan 2016


%\cite{Niemi:2015qia}
\bibitem{Niemi:2015qia} 
  H.~Niemi, K.~J.~Eskola and R.~Paatelainen,
  %``Event-by-event fluctuations in perturbative QCD + saturation + hydro model: pinning down QCD matter shear viscosity in ultrarelativistic heavy-ion collisions,''
  arXiv:1505.02677 [hep-ph].
  %%CITATION = ARXIV:1505.02677;%%
  %20 citations counted in INSPIRE as of 19 Jan 2016


%\cite{Habich:2014jna}
\bibitem{Habich:2014jna} 
  M.~Habich, J.~L.~Nagle and P.~Romatschke,
  %``Particle spectra and HBT radii for simulated central nuclear collisions of C + C, Al + Al, Cu + Cu, Au + Au, and Pb + Pb from $\sqrt{s}=62.4$ - $2760$ GeV,''
  Eur.\ Phys.\ J.\ C {\bf 75}, no. 1, 15 (2015)
  doi:10.1140/epjc/s10052-014-3206-7
  [arXiv:1409.0040 [nucl-th]].
  %%CITATION = doi:10.1140/epjc/s10052-014-3206-7;%%
  %13 citations counted in INSPIRE as of 19 Jan 2016


%\cite{Liu:2015nwa}
\bibitem{Liu:2015nwa} 
  J.~Liu, C.~Shen and U.~Heinz,
  %``Pre-equilibrium evolution effects on heavy-ion collision observables,''
  Phys.\ Rev.\ C {\bf 91}, no. 6, 064906 (2015)
  [Phys.\ Rev.\ C {\bf 92}, no. 4, 049904 (2015)]
  doi:10.1103/PhysRevC.92.049904, 10.1103/PhysRevC.91.064906
  [arXiv:1504.02160 [nucl-th]].
  %%CITATION = doi:10.1103/PhysRevC.92.049904, 10.1103/PhysRevC.91.064906;%%
  %4 citations counted in INSPIRE as of 19 Jan 2016


%\cite{Vredevoogd:2008id}
\bibitem{Vredevoogd:2008id} 
  J.~Vredevoogd and S.~Pratt,
  %``Universal Flow in the First Stage of Relativistic Heavy Ion Collisions,''
  Phys.\ Rev.\ C {\bf 79}, 044915 (2009)
  doi:10.1103/PhysRevC.79.044915
  [arXiv:0810.4325 [nucl-th]].
  %%CITATION = doi:10.1103/PhysRevC.79.044915;%%
  %59 citations counted in INSPIRE as of 19 Jan 2016


%\cite{vanderSchee:2012qj}
\bibitem{vanderSchee:2012qj} 
  W.~van der Schee,
  %``Holographic thermalization with radial flow,''
  Phys.\ Rev.\ D {\bf 87}, no. 6, 061901 (2013)
  doi:10.1103/PhysRevD.87.061901
  [arXiv:1211.2218 [hep-th]].
  %%CITATION = doi:10.1103/PhysRevD.87.061901;%%
  %28 citations counted in INSPIRE as of 19 Jan 2016


%\cite{Romatschke:2013re}
\bibitem{Romatschke:2013re} 
  P.~Romatschke and J.~D.~Hogg,
  %``Pre-Equilibrium Radial Flow from Central Shock-Wave Collisions in AdS5,''
  JHEP {\bf 1304}, 048 (2013)
  doi:10.1007/JHEP04(2013)048
  [arXiv:1301.2635 [hep-th]].
  %%CITATION = doi:10.1007/JHEP04(2013)048;%%
  %12 citations counted in INSPIRE as of 19 Jan 2016


%\cite{vanderSchee:2013pia}
\bibitem{vanderSchee:2013pia} 
  W.~van der Schee, P.~Romatschke and S.~Pratt,
  %``Fully Dynamical Simulation of Central Nuclear Collisions,''
  Phys.\ Rev.\ Lett.\  {\bf 111}, no. 22, 222302 (2013)
  doi:10.1103/PhysRevLett.111.222302
  [arXiv:1307.2539].
  %%CITATION = doi:10.1103/PhysRevLett.111.222302;%%
  %58 citations counted in INSPIRE as of 19 Jan 2016


%\cite{Luzum:2011mm}
\bibitem{Luzum:2011mm} 
  M.~Luzum,
  %``Flow fluctuations and long-range correlations: elliptic flow and beyond,''
  J.\ Phys.\ G {\bf 38}, 124026 (2011)
  doi:10.1088/0954-3899/38/12/124026
  [arXiv:1107.0592 [nucl-th]].
  %%CITATION = doi:10.1088/0954-3899/38/12/124026;%%
  %37 citations counted in INSPIRE as of 19 Jan 2016


%\cite{Gale:2013da}
\bibitem{Gale:2013da} 
  C.~Gale, S.~Jeon and B.~Schenke,
  %``Hydrodynamic Modeling of Heavy-Ion Collisions,''
  Int.\ J.\ Mod.\ Phys.\ A {\bf 28}, 1340011 (2013)
  doi:10.1142/S0217751X13400113
  [arXiv:1301.5893 [nucl-th]].
  %%CITATION = doi:10.1142/S0217751X13400113;%%
  %155 citations counted in INSPIRE as of 19 Jan 2016


%\cite{Bhalerao:2011yg}
\bibitem{Bhalerao:2011yg} 
  R.~S.~Bhalerao, M.~Luzum and J.~Y.~Ollitrault,
  %``Determining initial-state fluctuations from flow measurements in heavy-ion collisions,''
  Phys.\ Rev.\ C {\bf 84}, 034910 (2011)
  doi:10.1103/PhysRevC.84.034910
  [arXiv:1104.4740 [nucl-th]].
  %%CITATION = doi:10.1103/PhysRevC.84.034910;%%
  %91 citations counted in INSPIRE as of 19 Jan 2016


%\cite{Teaney:2012ke}
\bibitem{Teaney:2012ke} 
  D.~Teaney and L.~Yan,
  %``Non linearities in the harmonic spectrum of heavy ion collisions with ideal and viscous hydrodynamics,''
  Phys.\ Rev.\ C {\bf 86}, 044908 (2012)
  doi:10.1103/PhysRevC.86.044908
  [arXiv:1206.1905 [nucl-th]].
  %%CITATION = doi:10.1103/PhysRevC.86.044908;%%
  %78 citations counted in INSPIRE as of 19 Jan 2016


%\cite{Yan:2015jma}
\bibitem{Yan:2015jma} 
  L.~Yan and J.~Y.~Ollitrault,
  %``$\nu_4, \nu_5, \nu_6, \nu_7$: nonlinear hydrodynamic response versus LHC data,''
  Phys.\ Lett.\ B {\bf 744}, 82 (2015)
  doi:10.1016/j.physletb.2015.03.040
  [arXiv:1502.02502 [nucl-th]].
  %%CITATION = doi:10.1016/j.physletb.2015.03.040;%%
  %5 citations counted in INSPIRE as of 19 Jan 2016


%\cite{Yan:2014nsa}
\bibitem{Yan:2014nsa} 
  L.~Yan, J.~Y.~Ollitrault and A.~M.~Poskanzer,
  %``Azimuthal Anisotropy Distributions in High-Energy Collisions,''
  Phys.\ Lett.\ B {\bf 742}, 290 (2015)
  doi:10.1016/j.physletb.2015.01.039
  [arXiv:1408.0921 [nucl-th]].
  %%CITATION = doi:10.1016/j.physletb.2015.01.039;%%
  %9 citations counted in INSPIRE as of 19 Jan 2016


%\cite{Bjorken:1982qr}
\bibitem{Bjorken:1982qr} 
  J.~D.~Bjorken,
  %``Highly Relativistic Nucleus-Nucleus Collisions: The Central Rapidity Region,''
  Phys.\ Rev.\ D {\bf 27}, 140 (1983).
  doi:10.1103/PhysRevD.27.140
  %%CITATION = doi:10.1103/PhysRevD.27.140;%%
  %2407 citations counted in INSPIRE as of 19 Jan 2016


%\cite{Noronha-Hostler:2013gga}
\bibitem{Noronha-Hostler:2013gga} 
  J.~Noronha-Hostler, G.~S.~Denicol, J.~Noronha, R.~P.~G.~Andrade and F.~Grassi,
  %``Bulk Viscosity Effects in Event-by-Event Relativistic Hydrodynamics,''
  Phys.\ Rev.\ C {\bf 88}, 044916 (2013)
  doi:10.1103/PhysRevC.88.044916
  [arXiv:1305.1981 [nucl-th]].
  %%CITATION = doi:10.1103/PhysRevC.88.044916;%%
  %48 citations counted in INSPIRE as of 19 Jan 2016


%\cite{Noronha-Hostler:2014dqa}
\bibitem{Noronha-Hostler:2014dqa} 
  J.~Noronha-Hostler, J.~Noronha and F.~Grassi,
  %``Bulk viscosity-driven suppression of shear viscosity effects on the flow harmonics at energies available at the BNL Relativistic Heavy Ion Collider,''
  Phys.\ Rev.\ C {\bf 90}, no. 3, 034907 (2014)
  doi:10.1103/PhysRevC.90.034907
  [arXiv:1406.3333 [nucl-th]].
  %%CITATION = doi:10.1103/PhysRevC.90.034907;%%
  %26 citations counted in INSPIRE as of 19 Jan 2016


%\cite{Miller:2007ri}
\bibitem{Miller:2007ri} 
  M.~L.~Miller, K.~Reygers, S.~J.~Sanders and P.~Steinberg,
  %``Glauber modeling in high energy nuclear collisions,''
  Ann.\ Rev.\ Nucl.\ Part.\ Sci.\  {\bf 57}, 205 (2007)
  doi:10.1146/annurev.nucl.57.090506.123020
  [nucl-ex/0701025].
  %%CITATION = doi:10.1146/annurev.nucl.57.090506.123020;%%
  %685 citations counted in INSPIRE as of 19 Jan 2016


%\cite{Alver:2008aq}
\bibitem{Alver:2008aq} 
  B.~Alver, M.~Baker, C.~Loizides and P.~Steinberg,
  %``The PHOBOS Glauber Monte Carlo,''
  arXiv:0805.4411 [nucl-ex].
  %%CITATION = ARXIV:0805.4411;%%
  %194 citations counted in INSPIRE as of 19 Jan 2016


%\cite{Rybczynski:2013yba}
\bibitem{Rybczynski:2013yba} 
  M.~Rybczynski, G.~Stefanek, W.~Broniowski and P.~Bozek,
  %``GLISSANDO 2 : GLauber Initial-State Simulation AND mOre?, ver. 2,''
  Comput.\ Phys.\ Commun.\  {\bf 185}, 1759 (2014)
  doi:10.1016/j.cpc.2014.02.016
  [arXiv:1310.5475 [nucl-th]].
  %%CITATION = doi:10.1016/j.cpc.2014.02.016;%%
  %36 citations counted in INSPIRE as of 19 Jan 2016


%\cite{Hirano:2009ah}
\bibitem{Hirano:2009ah} 
  T.~Hirano and Y.~Nara,
  %``Eccentricity fluctuation effects on elliptic flow in relativistic heavy ion collisions,''
  Phys.\ Rev.\ C {\bf 79}, 064904 (2009)
  doi:10.1103/PhysRevC.79.064904
  [arXiv:0904.4080 [nucl-th]].
  %%CITATION = doi:10.1103/PhysRevC.79.064904;%%
  %91 citations counted in INSPIRE as of 19 Jan 2016


%\cite{Policastro:2001yc}
\bibitem{Policastro:2001yc} 
  G.~Policastro, D.~T.~Son and A.~O.~Starinets,
  %``The Shear viscosity of strongly coupled N=4 supersymmetric Yang-Mills plasma,''
  Phys.\ Rev.\ Lett.\  {\bf 87}, 081601 (2001)
  doi:10.1103/PhysRevLett.87.081601
  [hep-th/0104066].
  %%CITATION = doi:10.1103/PhysRevLett.87.081601;%%
  %1035 citations counted in INSPIRE as of 19 Jan 2016


%\cite{Kovtun:2004de}
\bibitem{Kovtun:2004de} 
  P.~Kovtun, D.~T.~Son and A.~O.~Starinets,
  %``Viscosity in strongly interacting quantum field theories from black hole physics,''
  Phys.\ Rev.\ Lett.\  {\bf 94}, 111601 (2005)
  doi:10.1103/PhysRevLett.94.111601
  [hep-th/0405231].
  %%CITATION = doi:10.1103/PhysRevLett.94.111601;%%
  %1568 citations counted in INSPIRE as of 19 Jan 2016


%\cite{NoronhaHostler:2008ju}
\bibitem{NoronhaHostler:2008ju} 
  J.~Noronha-Hostler, J.~Noronha and C.~Greiner,
  %``Transport Coefficients of Hadronic Matter near T(c),''
  Phys.\ Rev.\ Lett.\  {\bf 103}, 172302 (2009)
  doi:10.1103/PhysRevLett.103.172302
  [arXiv:0811.1571 [nucl-th]].
  %%CITATION = doi:10.1103/PhysRevLett.103.172302;%%
  %133 citations counted in INSPIRE as of 19 Jan 2016


%\cite{Niemi:2011ix}
\bibitem{Niemi:2011ix} 
  H.~Niemi, G.~S.~Denicol, P.~Huovinen, E.~Molnar and D.~H.~Rischke,
  %``Influence of the shear viscosity of the quark-gluon plasma on elliptic flow in ultrarelativistic heavy-ion collisions,''
  Phys.\ Rev.\ Lett.\  {\bf 106}, 212302 (2011)
  doi:10.1103/PhysRevLett.106.212302
  [arXiv:1101.2442 [nucl-th]].
  %%CITATION = doi:10.1103/PhysRevLett.106.212302;%%
  %151 citations counted in INSPIRE as of 19 Jan 2016


%\cite{Huovinen:2009yb}
\bibitem{Huovinen:2009yb} 
  P.~Huovinen and P.~Petreczky,
  %``QCD Equation of State and Hadron Resonance Gas,''
  Nucl.\ Phys.\ A {\bf 837}, 26 (2010)
  doi:10.1016/j.nuclphysa.2010.02.015
  [arXiv:0912.2541 [hep-ph]].
  %%CITATION = doi:10.1016/j.nuclphysa.2010.02.015;%%
  %289 citations counted in INSPIRE as of 19 Jan 2016


%\cite{Teaney:2003kp}
\bibitem{Teaney:2003kp} 
  D.~Teaney,
  %``The Effects of viscosity on spectra, elliptic flow, and HBT radii,''
  Phys.\ Rev.\ C {\bf 68}, 034913 (2003)
  doi:10.1103/PhysRevC.68.034913
  [nucl-th/0301099].
  %%CITATION = doi:10.1103/PhysRevC.68.034913;%%
  %599 citations counted in INSPIRE as of 19 Jan 2016


%\cite{Dusling:2009df}
\bibitem{Dusling:2009df} 
  K.~Dusling, G.~D.~Moore and D.~Teaney,
  %``Radiative energy loss and v(2) spectra for viscous hydrodynamics,''
  Phys.\ Rev.\ C {\bf 81}, 034907 (2010)
  doi:10.1103/PhysRevC.81.034907
  [arXiv:0909.0754 [nucl-th]].
  %%CITATION = doi:10.1103/PhysRevC.81.034907;%%
  %108 citations counted in INSPIRE as of 19 Jan 2016


%\cite{Chatrchyan:2012ta}
\bibitem{Chatrchyan:2012ta} 
  S.~Chatrchyan {\it et al.} [CMS Collaboration],
  %``Measurement of the elliptic anisotropy of charged particles produced in PbPb collisions at $\sqrt{s}_{NN}$=2.76 TeV,''
  Phys.\ Rev.\ C {\bf 87}, no. 1, 014902 (2013)
  doi:10.1103/PhysRevC.87.014902
  [arXiv:1204.1409 [nucl-ex]].
  %%CITATION = doi:10.1103/PhysRevC.87.014902;%%
  %146 citations counted in INSPIRE as of 19 Jan 2016


\bibitem{jackknife}
B.~Efron, 
%``Bootstrap Methods: Another Look at the Jackknife,''
Ann. Stat. {\bf 7}, 1 (1979).

%\cite{Drescher:2007cd}
\bibitem{Drescher:2007cd} 
  H.~J.~Drescher, A.~Dumitru, C.~Gombeaud and J.~Y.~Ollitrault,
  %``The Centrality dependence of elliptic flow, the hydrodynamic limit, and the viscosity of hot QCD,''
  Phys.\ Rev.\ C {\bf 76}, 024905 (2007)
  doi:10.1103/PhysRevC.76.024905
  [arXiv:0704.3553 [nucl-th]].
  %%CITATION = doi:10.1103/PhysRevC.76.024905;%%
  %190 citations counted in INSPIRE as of 19 Jan 2016


%\cite{Drescher:2006ca}
\bibitem{Drescher:2006ca} 
  H.-J.~Drescher and Y.~Nara,
  %``Effects of fluctuations on the initial eccentricity from the Color Glass Condensate in heavy ion collisions,''
  Phys.\ Rev.\ C {\bf 75}, 034905 (2007)
  doi:10.1103/PhysRevC.75.034905
  [nucl-th/0611017].
  %%CITATION = doi:10.1103/PhysRevC.75.034905;%%
  %155 citations counted in INSPIRE as of 19 Jan 2016


%\cite{Bhalerao:2011bp}
\bibitem{Bhalerao:2011bp} 
  R.~S.~Bhalerao, M.~Luzum and J.~Y.~Ollitrault,
  %``Understanding anisotropy generated by fluctuations in heavy-ion collisions,''
  Phys.\ Rev.\ C {\bf 84}, 054901 (2011)
  doi:10.1103/PhysRevC.84.054901
  [arXiv:1107.5485 [nucl-th]].
  %%CITATION = doi:10.1103/PhysRevC.84.054901;%%
  %58 citations counted in INSPIRE as of 19 Jan 2016


%\cite{Blaizot:2014nia}
\bibitem{Blaizot:2014nia} 
  J.~P.~Blaizot, W.~Broniowski and J.~Y.~Ollitrault,
  %``Continuous description of fluctuating eccentricities,''
  Phys.\ Lett.\ B {\bf 738}, 166 (2014)
  doi:10.1016/j.physletb.2014.09.028
  [arXiv:1405.3572 [nucl-th]].
  %%CITATION = doi:10.1016/j.physletb.2014.09.028;%%
  %2 citations counted in INSPIRE as of 19 Jan 2016


%\cite{Luzum:2008cw}
\bibitem{Luzum:2008cw} 
  M.~Luzum and P.~Romatschke,
  %``Conformal Relativistic Viscous Hydrodynamics: Applications to RHIC results at s(NN)**(1/2) = 200-GeV,''
  Phys.\ Rev.\ C {\bf 78}, 034915 (2008)
  [Phys.\ Rev.\ C {\bf 79}, 039903 (2009)]
  doi:10.1103/PhysRevC.78.034915, 10.1103/PhysRevC.79.039903
  [arXiv:0804.4015 [nucl-th]].
  %%CITATION = doi:10.1103/PhysRevC.78.034915, 10.1103/PhysRevC.79.039903;%%
  %577 citations counted in INSPIRE as of 19 Jan 2016


%\cite{Laine:2006cp}
\bibitem{Laine:2006cp} 
  M.~Laine and Y.~Schroder,
  %``Quark mass thresholds in QCD thermodynamics,''
  Phys.\ Rev.\ D {\bf 73}, 085009 (2006)
  doi:10.1103/PhysRevD.73.085009
  [hep-ph/0603048].
  %%CITATION = doi:10.1103/PhysRevD.73.085009;%%
  %219 citations counted in INSPIRE as of 19 Jan 2016


%\cite{Aguiar:2000hw}
\bibitem{Aguiar:2000hw} 
  C.~E.~Aguiar, T.~Kodama, T.~Osada and Y.~Hama,
  %``Smoothed particle hydrodynamics for relativistic heavy ion collisions,''
  J.\ Phys.\ G {\bf 27}, 75 (2001)
  doi:10.1088/0954-3899/27/1/306
  [hep-ph/0006239].
  %%CITATION = doi:10.1088/0954-3899/27/1/306;%%
  %105 citations counted in INSPIRE as of 19 Jan 2016


%\cite{Hama:2004rr}
\bibitem{Hama:2004rr} 
  Y.~Hama, T.~Kodama and O.~Socolowski, Jr.,
  %``Topics on hydrodynamic model of nucleus-nucleus collisions,''
  Braz.\ J.\ Phys.\  {\bf 35}, 24 (2005)
  doi:10.1590/S0103-97332005000100003
  [hep-ph/0407264].
  %%CITATION = doi:10.1590/S0103-97332005000100003;%%
  %126 citations counted in INSPIRE as of 19 Jan 2016


%\cite{Marrochio:2013wla}
\bibitem{Marrochio:2013wla} 
  H.~Marrochio, J.~Noronha, G.~S.~Denicol, M.~Luzum, S.~Jeon and C.~Gale,
  %``Solutions of Conformal Israel-Stewart Relativistic Viscous Fluid Dynamics,''
  Phys.\ Rev.\ C {\bf 91}, no. 1, 014903 (2015)
  doi:10.1103/PhysRevC.91.014903
  [arXiv:1307.6130 [nucl-th]].
  %%CITATION = doi:10.1103/PhysRevC.91.014903;%%
  %30 citations counted in INSPIRE as of 19 Jan 2016


%\cite{Schenke:2012wb}
\bibitem{Schenke:2012wb} 
  B.~Schenke, P.~Tribedy and R.~Venugopalan,
  %``Fluctuating Glasma initial conditions and flow in heavy ion collisions,''
  Phys.\ Rev.\ Lett.\  {\bf 108}, 252301 (2012)
  doi:10.1103/PhysRevLett.108.252301
  [arXiv:1202.6646 [nucl-th]].
  %%CITATION = doi:10.1103/PhysRevLett.108.252301;%%
  %184 citations counted in INSPIRE as of 19 Jan 2016


%\cite{Aad:2014vba}
\bibitem{Aad:2014vba} 
  G.~Aad {\it et al.} [ATLAS Collaboration],
  %``Measurement of flow harmonics with multi-particle cumulants in Pb+Pb collisions at $\sqrt{s_{\mathrm {NN}}}=2.76$  TeV with the ATLAS detector,''
  Eur.\ Phys.\ J.\ C {\bf 74}, no. 11, 3157 (2014)
  doi:10.1140/epjc/s10052-014-3157-z
  [arXiv:1408.4342 [hep-ex]].
  %%CITATION = doi:10.1140/epjc/s10052-014-3157-z;%%
  %14 citations counted in INSPIRE as of 19 Jan 2016


%\cite{Borghini:2001vi}
\bibitem{Borghini:2001vi} 
  N.~Borghini, P.~M.~Dinh and J.~Y.~Ollitrault,
  %``Flow analysis from multiparticle azimuthal correlations,''
  Phys.\ Rev.\ C {\bf 64}, 054901 (2001)
  doi:10.1103/PhysRevC.64.054901
  [nucl-th/0105040].
  %%CITATION = doi:10.1103/PhysRevC.64.054901;%%
  %280 citations counted in INSPIRE as of 19 Jan 2016


%\cite{Bilandzic:2010jr}
\bibitem{Bilandzic:2010jr} 
  A.~Bilandzic, R.~Snellings and S.~Voloshin,
  %``Flow analysis with cumulants: Direct calculations,''
  Phys.\ Rev.\ C {\bf 83}, 044913 (2011)
  doi:10.1103/PhysRevC.83.044913
  [arXiv:1010.0233 [nucl-ex]].
  %%CITATION = doi:10.1103/PhysRevC.83.044913;%%
  %125 citations counted in INSPIRE as of 19 Jan 2016


%\cite{Bhalerao:2014xra}
\bibitem{Bhalerao:2014xra} 
  R.~S.~Bhalerao, J.~Y.~Ollitrault and S.~Pal,
  %``Characterizing flow fluctuations with moments,''
  Phys.\ Lett.\ B {\bf 742}, 94 (2015)
  doi:10.1016/j.physletb.2015.01.019
  [arXiv:1411.5160 [nucl-th]].
  %%CITATION = doi:10.1016/j.physletb.2015.01.019;%%
  %9 citations counted in INSPIRE as of 19 Jan 2016


%\cite{Alver:2010dn}
\bibitem{Alver:2010dn} 
  B.~H.~Alver, C.~Gombeaud, M.~Luzum and J.~Y.~Ollitrault,
  %``Triangular flow in hydrodynamics and transport theory,''
  Phys.\ Rev.\ C {\bf 82}, 034913 (2010)
  doi:10.1103/PhysRevC.82.034913
  [arXiv:1007.5469 [nucl-th]].
  %%CITATION = doi:10.1103/PhysRevC.82.034913;%%
  %240 citations counted in INSPIRE as of 19 Jan 2016


%\cite{Aad:2013xma}
\bibitem{Aad:2013xma} 
  G.~Aad {\it et al.} [ATLAS Collaboration],
  %``Measurement of the distributions of event-by-event flow harmonics in lead-lead collisions at = 2.76 TeV with the ATLAS detector at the LHC,''
  JHEP {\bf 1311}, 183 (2013)
  doi:10.1007/JHEP11(2013)183
  [arXiv:1305.2942 [hep-ex]].
  %%CITATION = doi:10.1007/JHEP11(2013)183;%%
  %111 citations counted in INSPIRE as of 19 Jan 2016


%\cite{Retinskaya:2013gca}
\bibitem{Retinskaya:2013gca} 
  E.~Retinskaya, M.~Luzum and J.~Y.~Ollitrault,
  %``Constraining models of initial conditions with elliptic and triangular flow data,''
  Phys.\ Rev.\ C {\bf 89}, no. 1, 014902 (2014)
  doi:10.1103/PhysRevC.89.014902
  [arXiv:1311.5339 [nucl-th]].
  %%CITATION = doi:10.1103/PhysRevC.89.014902;%%
  %26 citations counted in INSPIRE as of 19 Jan 2016


%\cite{Renk:2014jja}
\bibitem{Renk:2014jja} 
  T.~Renk and H.~Niemi,
  %``Constraints from $v_2$ fluctuations for the initial-state geometry of heavy-ion collisions,''
  Phys.\ Rev.\ C {\bf 89}, no. 6, 064907 (2014)
  doi:10.1103/PhysRevC.89.064907
  [arXiv:1401.2069 [nucl-th]].
  %%CITATION = doi:10.1103/PhysRevC.89.064907;%%
  %17 citations counted in INSPIRE as of 19 Jan 2016


%\cite{Rybczynski:2015wva}
\bibitem{Rybczynski:2015wva} 
  M.~Rybczynski and W.~Broniowski,
  %``Fluctuations of flow harmonics in Pb+Pb collisions at $\sqrt{s_{NN}}=2.76$~TeV from the Glauber model,''
  arXiv:1510.08242 [nucl-th].
  %%CITATION = ARXIV:1510.08242;%%
  %3 citations counted in INSPIRE as of 19 Jan 2016


%\cite{Noronha-Hostler:2015coa}
\bibitem{Noronha-Hostler:2015coa} 
  J.~Noronha-Hostler, J.~Noronha and M.~Gyulassy,
  %``Sensitivity of flow harmonics to sub-nucleon scale fluctuations in heavy ion collisions,''
  arXiv:1508.02455 [nucl-th].
  %%CITATION = ARXIV:1508.02455;%%
  %6 citations counted in INSPIRE as of 19 Jan 2016

 
\end{thebibliography}
\end{document}